\documentclass[finallayout,an,fleqn]{w-art}
\usepackage{times}
\usepackage{w-thm}
\theoremstyle{plain}

\theoremstyle{definition}

\usepackage[]{graphicx}
\chardef\bslash=`\\ 

\begin{document}
\DOIsuffix{theDOIsuffix}
\Volume{324}
\Issue{S1}
\Copyrightissue{S1}
\Month{01}
\Year{2003}
\pagespan{1}{}
\Receiveddate{15 November 2002}
\Reviseddate{30 November 2002}
\Accepteddate{2 December 2002}
\Dateposted{3 December 2002}



\title[Gas physics and dynamics in the central 50~pc of the Galaxy]{Gas physics and dynamics in the central 50~pc of the Galaxy}


\author[B.~Vollmer]{B.~Vollmer\footnote{Corresponding
     author: e-mail: {\sf bvollmer@mpifr-bonn.mpg.de}, Phone: +00\,49\,228\,525315,
     Fax: +00\,49\,228\,525436}\inst{1}} \address[\inst{1}]{Max-Planck-Institut f\"{u}r Radioastronomie, Auf dem H\"{u}gel 69, 53121 Bonn, Germany}
\author[W.J.~Duschl]{W.J.~Duschl\footnote{email: wjd@ita.uni-heidelberg.de}\inst{2,1}}
\address[\inst{2}]{Institut f\"ur Theoretische
	      Astrophysik der Universit\"at Heidelberg, Tiergartenstra{\ss}e 15,
              69121 Heidelberg, Germany}
\author[R. Zylka]{R. Zylka\footnote{email: zylka@iram.fr}\inst{3}}
\address[\inst{3}]{IRAM, 300, rue de la piscine, 38406 Saint Martin d'H\`eres, France}

\begin{abstract}
We present models the gas physics and dynamics of the inner 50~pc of the Galaxy. 
In a first step the gas properties of an isolated clumpy circumnuclear disk were analytically
investigated. We took the external UV radiation field, the gravitational potential,
and the observed gas temperature into account. The model includes a description
of the properties of individual gas clumps on small scales, and a treatment
of the circumnuclear disk as a quasi-continuous accretion disk on large scales.
In a second step the dynamics of an isolated circumnuclear disk were investigated
with the help of a collisional N-body code. The environment of the disk is taken
into account in a third step, where we calculated a pro- and a retrograde encounter
of an infalling gas cloud with a pre-existing circumnuclear disk. In order to 
constrain the dynamical model, we used the NIR absorption of the giant molecular clouds
located within the inner 50~pc of the Galaxy to reconstruct their line-of-sight
distribution.    
\end{abstract}
\maketitle                   

\section{Introduction}

During the discussion led by R.~Narayan at this conference it became clear
that the mass accretion rate onto the central black hole in the Galactic
Centre at a radius of $<$1~pc is $\sim 10^{-8}$~M$_{\odot}$\,yr$^{-1}$.
The central black hole is thus extremely sub-Eddington. In order to
understand the fueling mechanisms of the central engine, there is an
inevitable need for understanding the gas physics and dynamics in the
inner $\sim$50~pc of the Galaxy. Gas that flows radially into the Galactic
Centre has to pass several barriers. At large scales ($<$kpc) the gas has
to cross the resonance of the inner Lindblad radius. According to the
gravitational potential of the Galactic Bulge region, there might be a
second inner Lindblad radius that the gas has to overcome. When the gas
finally arrives in the inner 200~pc of the Galaxy, it is very clumpy and
has a volume filling factor of a few percent (Launhardt et al. 2002). In
this environment five different environmental effects determine the
structure of the ISM: (i) the stellar radiation field, (ii) stellar winds,
(iii) the shear due to differential rotation, (iv) instabilities due to
self-gravitation, and (v) supernovae. We modeled analytically the
properties of the gas located in the inner 20~pc including the effects
(i)--(iv) (Vollmer \& Duschl 2001a, 2001b). In a second step we
investigated the collision of an external gas cloud falling onto an
existing disk structure in the Galactic Centre (Vollmer \& Duschl 2002).
Finally, the line-of-sight distribution of the giant molecular clouds in
the inner 50~pc of the Galaxy were reconstructed using their NIR
absoprtion (Vollmer et al. 2003).

\section{The analytical model (Vollmer \& Duschl 2001a, 2001b) \label{sec:analytical}}

\begin{figure}[htb]
\begin{center}
\includegraphics[width=8cm, height=!]{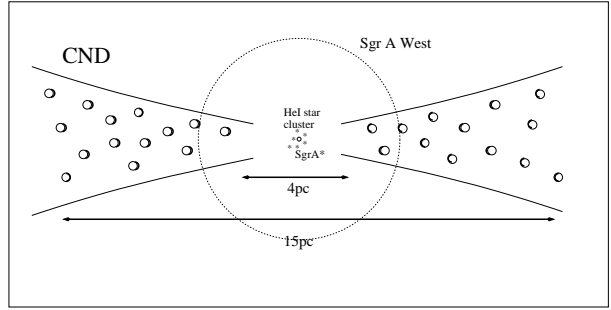}
\caption{Sketch of the inner 15~pc of the Galaxy. The Circumnuclear Disk (CND)
consists of small, dense gas clumps that are illuminated by the central He{\sc i}
star cluster. The dotted line delineates approximately the extension of the
H{\sc ii} region Sgr~A~West.}
\label{fig:vollmerb_fig1}
\end{center}
\end{figure}

In the inner 20~pc of the galaxy the gas is very clumpy with a volume filling
factor of $\sim$1\%. These clumps have masses of $\sim$30~M$_{\odot}$ and sizes
of $\sim$0.1~pc (see e.g. Jackson et al. 1993). They are illuminated by the UV
radiation field of the central He{\sc i} star cluster and form a ring-like structure
that is known as the Circumnuclear Disk (CND) (G\"{u}sten et al. 1987).
Fig.~\ref{fig:vollmerb_fig1} illustrates the situation. The whole CND has a total
gas mass of several 10$^{4}$~M$_{\odot}$. The inner edge is located at a radius
of $\sim$2~pc where the density of the neutral gas drops by more than an order
of magnitude. The gas in the central 4~pc is ionized and forms the giant
H{\sc ii} region Sgr~A~West.

\begin{figure}[htb]
\begin{center}
\includegraphics[width=6cm, height=!]{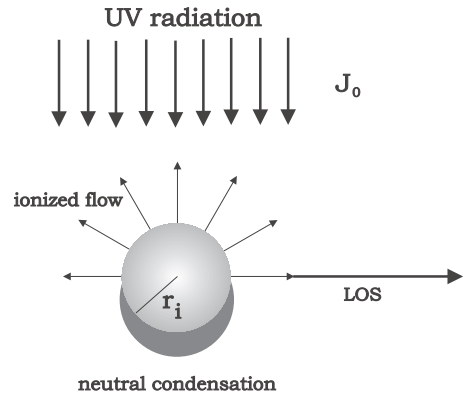}
\caption{Illustration of the partially ionized globule model described in the text.
	The UV radiation comes from the direction of the Galactic Centre.
	The boundary opposite to the Galactic Centre is determined by the \
	gas pressure of the surrounding ionized low density gas.}
\label{fig:vollmerb_fig2}
\end{center}
\end{figure}

Our analytical model to describe the gas properties consists of two parts:\\
(i) Small scale: the gas clumps are described as isothermal spheres that are partially 
ionized by the UV radiation field (Fig.~\ref{fig:vollmerb_fig2}). 
The gas temperature of the clouds is assumed to be proportional to $R^{-\frac{1}{2}}$.\\
(ii) Large scale: the properties of clouds located at different distances from the Galactic
Centre are determined. The clumpy gas distribution is then smeared out to obtain
a quasi-continuous disk structure that is described by the standard set of 
accretion disk equations with a modified viscosity prescription taking the
radiative energy dissipation during clump--clump collisions into account.

Both models are connected via the central density of the clumps that is 
proportional to the central density of the disk: 
$\rho_{\rm cl} = \Phi_{\rm V}^{-1}\,\rho_{\rm disk}$, where $\Phi_{\rm V}=0.01$
is the volume filling factor. In addition, the clump properties influence the
disk viscosity via the local energy dissipation rate. 
The main results of this modelling are:\\
(i) There are two solutions for our set of equations that correspond to
two stable clump regimes: the observed heavy clumps $\sim$10~M$_{\odot}$
and the stripped cores of the heavy clouds with masses between
10$^{-5}$ and 10$^{-4}$\,M$_{\odot}$.\\
(ii) Within the disk, the number of collisions between clouds is very
low ($\overline{n}_{\rm coll} \sim 10^{-4}$ yr$^{-1}$ for about
500 clouds).\\
(iii) The inferred mass accretion rate for the isolated CND is 
$\dot{M} \simeq 10^{-4}$\,M$_{\odot}$\,yr$^{-1}$.\\
(iv) The CND is much more stable and has a much longer lifetime
($\sim 10^{7}$ yr) than previously assumed.

\begin{figure}[htb] 
\begin{center}
\includegraphics[width=8cm, height=!]{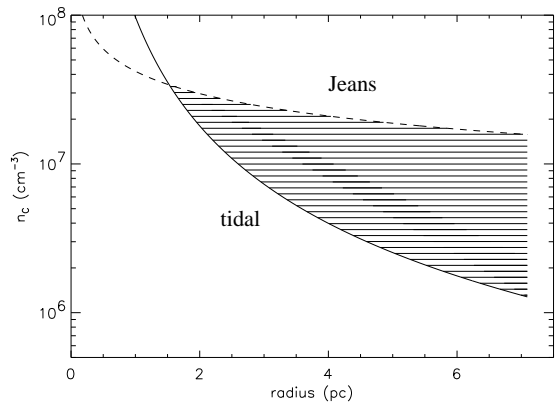} 
\caption{Centraldensity of the heavy clouds versus the
      distance to the Galactic Centre. Dashed line: maximum central
      density above which gravitational collapse occurs. Solid
      line: minimum density in order to resist tidal shear. Dashed
      surface: range of densities where clouds are gravitationally
      and tidally stable.}
\label{fig:vollmerb_fig3}
\end{center}
\end{figure}

The disk clouds are exposed to strong stellar winds, the central UV
radiation field, and strong tidal shear. We show that stellar winds shape
the clouds, but do not affect their other properties. In order to resist
the strong shear due to differential rotation each cloud's central
densities must increase with decreasing distance to the Galactic
Centre. Its
outer radius is determined by the UV radiation field at the illuminated
side and by the external gas pressure at the shadowed side opposite to the
direction of the Galactic Centre. We show analytically that for a
temperature gradient of the form $T \propto R^{-\frac{1}{2}}$ the radius
of the massive clouds are constant irrespective of their location in the
disk. Thus, clouds that can resist tidal shear will become more and more
massive when approaching the Galactic Centre. Finally, they will collapse
when their masses exceed the Jeans limit. We suggest that the cloud
distribution within the CND reflects two selection effects. First, the
clouds have to be dense enough to resist tidal shear and second, the
clouds that are too massive collapse and will form stars. Magnetic fields
and rotation stabilize otherwise gravitationally unstable clouds.  This
mechanism naturally explains the existence of the inner edge of the CND:
Fig.~\ref{fig:vollmerb_fig3} shows the central density of the heavy clouds
versus the distance from the Galactic Centre. The limits for gravitational
collapse and tidal disruption cross at $R \sim$2~pc. The dashed surface
represents the range of densities where clouds are gravitationally and
tidally stable. 

\section{The numerical model (Vollmer \& Duschl 2002)}

In a next step, we use our knowledge acquired with the analytical model to
build a realistic dynamical model in order to investigate the gas dynamics
in the inner 50~pc of the Galaxy. Since a circumnuclear disk (CND)
consisting of small clumps with a tiny volume-filling factor can be
long-lived (several Myr, see Sect.~\ref{sec:analytical}), we study the
scenario where a part of a giant molecular cloud falls onto a pre-existing
CND. We use a collisional N-body code where each particle represents a gas
clump with a certain mass and radius. The infalling gas cloud also
consists of a number of small subclumps. These clumps have masses around
the observed value of 30~M$_{\odot}$. When a clump approaches the Galactic
Centre closer than 2~pc, it is assumed to be destroyed by tidal forces or
by gravitational collapse (see Sect.~\ref{sec:analytical}). These clumps
are counted as accreted.

\begin{figure}
\begin{center}
\includegraphics[width=\textwidth, height=!]{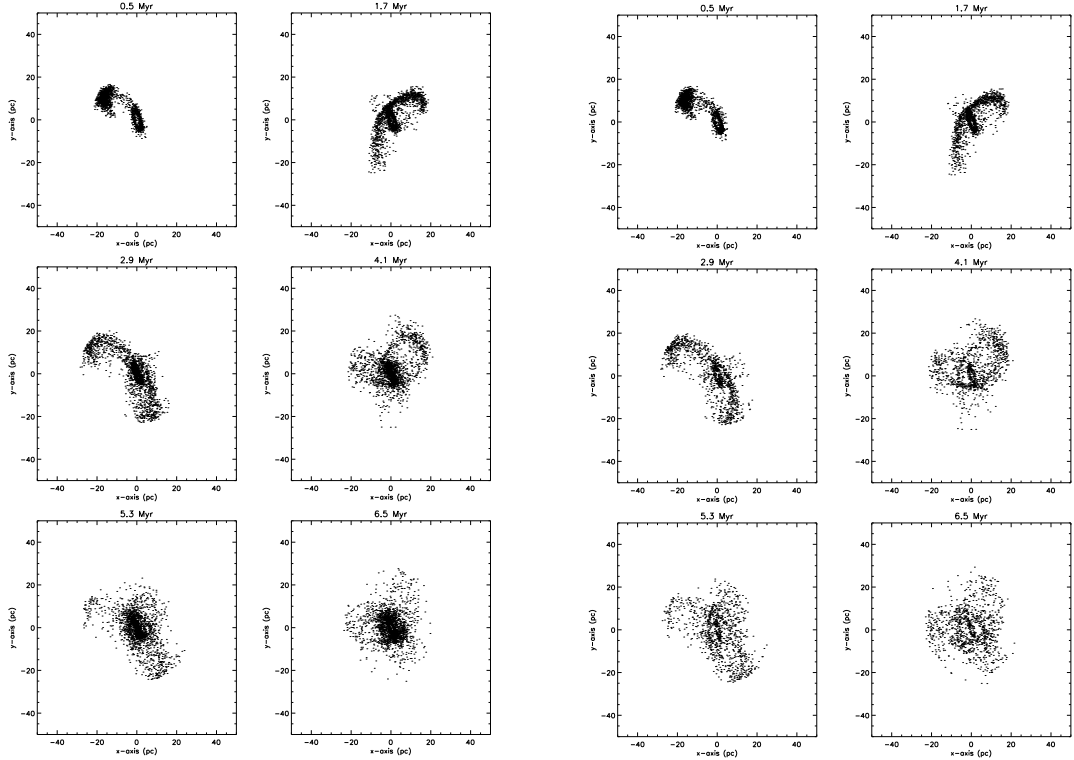}
\caption{The evolution of the cloud infall into the
	Galactic Centre as the observer would see it from the Earth.
	Left side: prograde encounter. Right side: retrograde encounter.
	The elapsed time is plotted on the top of each frame.}
\label{fig:vollmerb_fig4}
\end{center}
\end{figure}

The collisional N-body code yields a realistic cloud collision rate that
depends on the cloud radius, density, and dispersion velocity. A realistic
simulation of an isolated CND shows the observed disk structure. This
simulation yields a mean collision time scale of a cloud of $t_{\rm coll}
\sim 2$~Myr and a mass accretion rate of $10^{-4}$~M$_{\odot}$\,yr$^{-1}
\leq \dot{M} \leq 10^{-3}$~M$_{\odot}$\,yr$^{-1}$.  The infalling cloud,
which has a mass of several 10$^{4}$~M$_{\odot}$, is assumed to be on (i)
a prograde orbit and (ii) a retrograde orbit with respect to this CND. We
study the resulting cloud--cloud collision rate and the mass accretion
rate using different loss rates of the kinetic energy during a collision.
Fig.~\ref{fig:vollmerb_fig4} shows a prograde (left side) and a retrograde
(right side) encounter with a loss of 10\% of the kinetic energy of the
clumps during a collision.
The main difference between the two simulations is that the CND is
destroyed 3-4~Myr after the first encounter with the external cloud in the
case of a retrograde encounter. Its mass is mainly accreted onto the
Galactic Centre. At the end of the simulation a second CND has formed that
has approximately the angular momentum of the infalling cloud, but there
is still a counter-rotating core visible, which represents the remnant of
the former CND. In the case of a prograde encounter, the infalling mass is
partly added to the pre-existing one. The outcome of this simulation is a
warped CND that is more massive than the pre-existing CND.  Within our
scenario of an encounter between an infalling molecular cloud and a CND,
it is possible that the observed He{\sc ii} star cluster in the Galactic
Centre has been formed by a retrograde encounter of a cloud with the CND
$\sim$7~Myr ago. The cloud that formed the He{\sc ii} star cluster has
been destroyed by tidal forces and can presently no longer be
distinguished as a single kinematical entity.

\section{The LOS distribution of the GMCs (Vollmer et al. 2003)}

\begin{figure}[htb]
\begin{center}
\includegraphics[width=8cm, height=!]{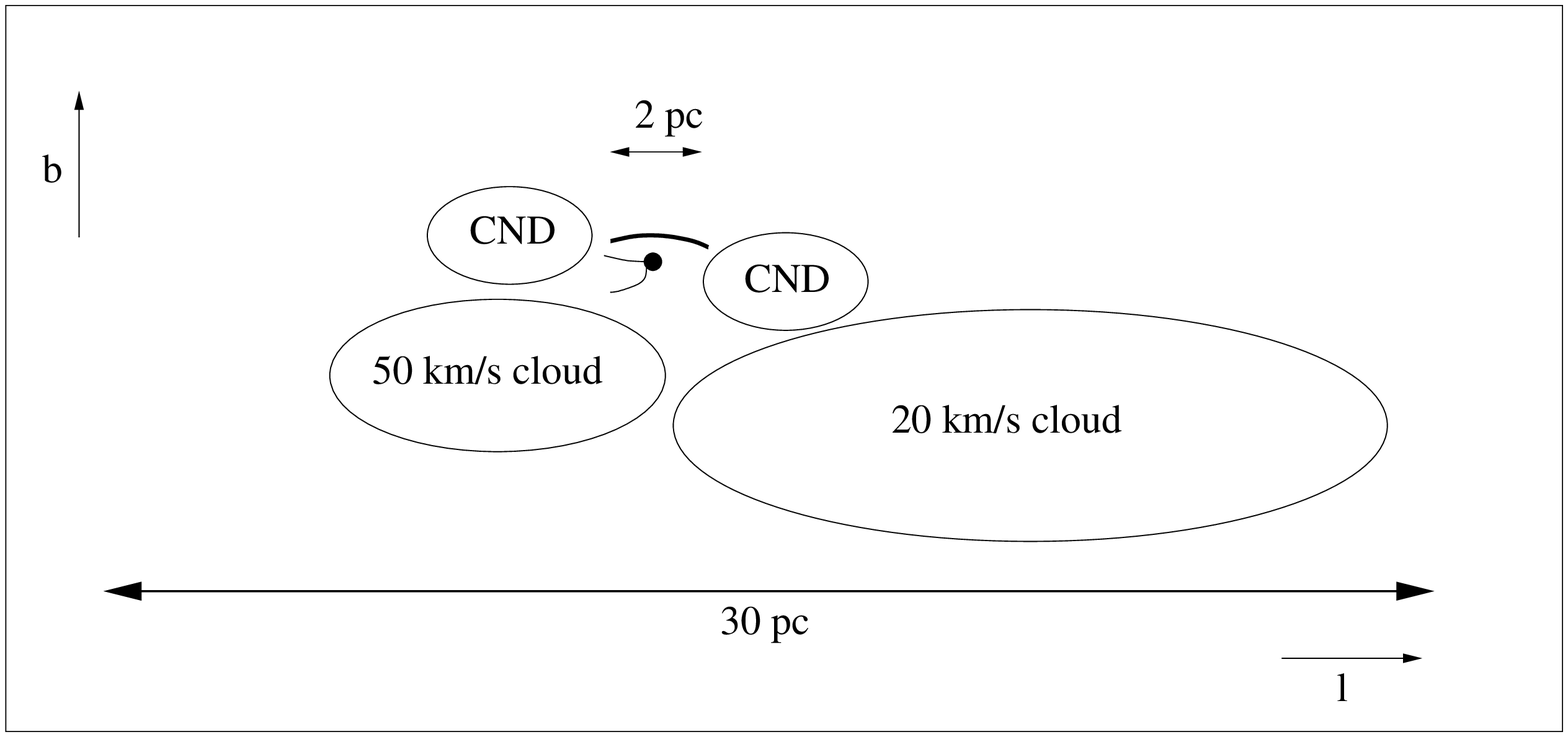}
\caption{Sketch of the inner 30~pc of the Galaxy. The central black dot represents the
He{\sc i} star cluster surrounded by the Minispiral and the CND (where only the
high velocity lobes are shown). Most of the mass is located at negative $b$.}
\label{fig:vollmerb_fig5}
\end{center}
\end{figure}

The fueling of the central black hole in the future depends mainly on the
present dynamics of the gas in the inner 50~pc of the Galaxy. Within this
region the gas is heavily clumped and mainly in the form of giant
molecular clouds.  Following Zylka et al. (1990) three main giant
molecular cloud complexes can be distinguished:
(i) Sgr~A East Core, a compact giant molecular cloud with a gas mass of
several 10$^{5}$ M$_{\odot}$ located north--east of Sgr~A$^{*}$.
(ii) The giant molecular cloud M-0.02-0.07 located to the east of
Sgr~A$^{*}$. Since its mean radial velocity is $\sim$50~km\,s$^{-1}$, it
is also called the 50~km\,s$^{-1}$ cloud. (iii) The GMC complex
M-0.13-0.08 located south of Sgr~A$^{*}$. Since its mean radial velocity 
is $\sim$20~km\,s$^{-1}$, it is also called the 20~km\,s$^{-1}$ cloud.
The Sgr A East core is part of the 50~km\,s$^{-1}$ cloud complex, thus we
will treat these features as a single structure. 
Fig.~\ref{fig:vollmerb_fig5} shows a sketch of the inner 30~pc of the
Galaxy, where the main features are indicated (Minispiral, CND,
20~km\,s$^{-1}$ cloud, 50~km\,s$^{-1}$ cloud).
If one wants to understand the gas dynamics in the central 50~pc of the
Galaxy, it is of crucial importance to know the line-of-sight distribution
of the giant molecular clouds (20 and 50~km\,s$^{-1}$ clouds) located in
this region.

\begin{figure}
\begin{center}
\includegraphics[width=10cm,height=!,angle=-90]{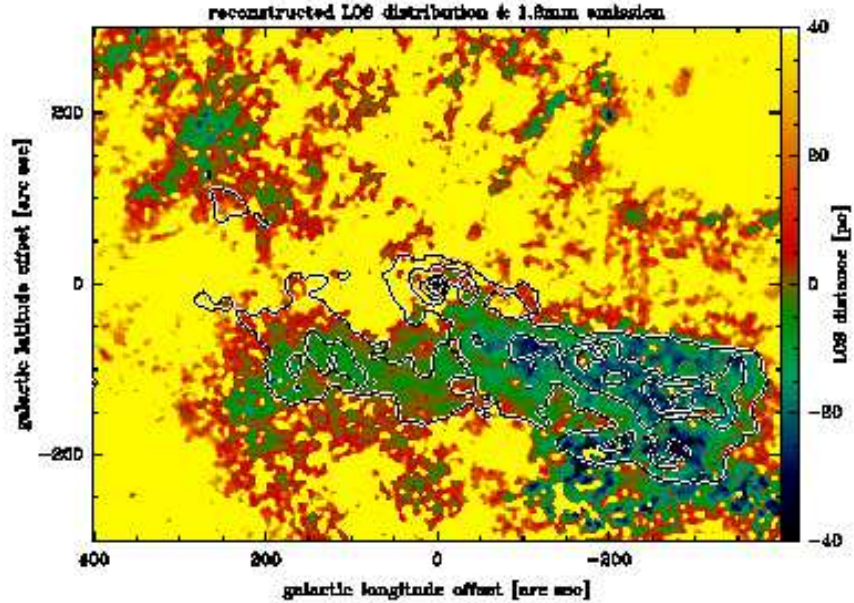}
\caption{Contours: IRAM 30m 1.2~mm observations of Zylka et al. (1998).
	Colors: LOS distance distribution filtered with a median filter 
of 11 pixel size.}
\label{fig:vollmerb_fig6}
\end{center}
\end{figure}

We reconstruct the line-of-sight distribution assuming (i) an
axis-symmetric stellar distribution and (ii) that the clouds are optically
thick and have an area filling factor $\sim$1, i.e. that they entirely
block the light from the stars located behind them.
Fig.~\ref{fig:vollmerb_fig6} shows the reconstructed LOS distribution of
the giant molecular clouds in colors. The IRAM 30m 1.2~mm observations of
Zylka et al. (1998) are overlayed as contours. Due to the method of
reconstruction, LOS distances close to Sgr~A$^{*}$ ($<$-10~pc) have a
small uncertainty, whereas larger LOS distances might be located up to a
factor 2 farther away from Sgr~A$^{*}$. The relative distances are robust
results. We found that:
\\ All structures seen in the 1.2~mm observations
(Zylka et al. 1998) and CS(2-1) observations (G\"{u}sten et al. in prep.)
are present in absorption.
\\ The 50~km\,s$^{-1}$ cloud complex is located
between 0~pc and -5~pc, i.e. in front of Sgr~A$^{*}$. It has a small LOS
distance gradient.
\\ The 20~km\,s$^{-1}$ cloud complex is located in front
of the 50~km\,s$^{-1}$ cloud complex. The subclump of strongest absorption
has a LOS distance between -40~pc and -20~pc.
\\ The CND is not seen in
absorption. This gives an upper limit to the cloud sizes within the CND of
$\sim$0.06~pc.

The combination of the LOS distribution of the gas and its kinematics will 
help to unravel the dynamics of the gas in the inner 50~pc of the Galaxy,
which represents the future fueling of the central black hole.

\section{Outlook}

From the present modeling of the gas in the inner 50~pc of the Galaxy we
have learned about crucial aspects of the gas physics and dynamics. We
found that the mass accretion rate into the central parsec is highly
variable and a period of almost no mass accretion is conceivable. This
period of starvation might last about 10$^{4}$~yr, which is the
cloud--cloud collision time within the CND.

We are now in the position to use the acquired knowledge to determine the
temporal behaviour of the mass accretion rate in the past and how it might
change in the future. Remaining questions are:\\ Can we find clear signs
of past interactions of an external cloud with a pre-existing
circumnuclear disk? Will the accreted mass exclusively come from the CND
in the near future? Will there be a major accretion event in the near
future, when a part of a massive giant molecular cloud interacts with the
CND? We already have the keys in our hand to unravel the exciting history
and future of the gas dynamics in the Galactic Centre.

\begin{acknowledgement}
This publication makes use of data products from the Two Micron All Sky
Survey, which is a joint project of the University of Massachusetts and
the Infrared Processing and Analysis Center/California Institute of
Technology, funded by the National Aeronautics and Space Administration
and the National Science Foundation.\\ BV thanks the {\it Deutsche
Forschungsgemeinschaft} for travel funding. \end{acknowledgement}

\end{document}